\documentclass[10pt,journal,a4paper]{IEEEtran}
\usepackage{amssymb,amsmath}
\usepackage{cuted}
\usepackage{flushend}
\usepackage{algorithm}
\usepackage{scalerel}
\usepackage{dsfont}
\usepackage[table]{xcolor}
\usepackage{gensymb}
\usepackage{mathbbol}
\usepackage{bbm}
\usepackage{algpseudocode}
\usepackage{cite}
\usepackage{graphicx,subfigure}
\usepackage{multirow}
\usepackage{psfrag}
\usepackage{url}
\usepackage{xcolor}
\usepackage[absolute,overlay]{textpos}
\usepackage{algpseudocode}
\usepackage{float}
\usepackage{lettrine}
\usepackage{indentfirst}
\usepackage{amsthm}

\newtheorem{theorem}{Theorem}

\usepackage{array}
\usepackage{psfrag}
\usepackage{url}
\usepackage{caption}
\usepackage[caption=false]{subfig}
\usepackage[font=footnotesize]{caption}
\usepackage{subfigure}

\makeatletter
\def\thickhline{%
  \noalign{\ifnum0=`}\fi\hrule \@height \thickarrayrulewidth \futurelet
   \reserved@a\@xthickhline}
\def\@xthickhline{\ifx\reserved@a\thickhline
               \vskip\doublerulesep
               \vskip-\thickarrayrulewidth
             \fi
      \ifnum0=`{\fi}}
\makeatother

\newlength{\thickarrayrulewidth}
\begin{document}
	
	\title{Extreme Value Theory-based Robust Minimum-Power Precoding for URLLC
}
	\author{
        \IEEEauthorblockN{Dian Echevarría Pérez, \IEEEmembership{Graduate Student Member, IEEE} \IEEEmembership{} %
        Onel L. Alcaraz López, \IEEEmembership{Member, IEEE}, Hirley Alves, \IEEEmembership{Member, IEEE}
        } 
        
		\thanks{The authors are with the Centre for Wireless Communications (CWC), University of Oulu, Finland. \{dian.echevarriaperez,  onel.alcarazlopez, hirley.alves\}@oulu.fi}

        \thanks{This research has been financially supported by Academy of Finland, 6G Flagship programme  (Grant no. 346208), and the Finnish Foundation for Technology Promotion.}
    }  
    \maketitle
	\begin{abstract}
     Channel state information (CSI) is crucial for
achieving ultra-reliable low-latency communication (URLLC)
in wireless networks. The main associated problems are the
CSI acquisition time, which impacts the delay requirements of
time-critical applications, and the estimation accuracy, which
degrades the signal-to-interference-plus-noise ratio (SINR), thus,
reducing reliability. In this work, we formulate and solve a
minimum-power precoding design problem simultaneously
serving multiple URLLC users in the downlink with imperfect
CSI availability. Specifically, we develop an algorithm that
exploits state-of-the-art precoding schemes such as the maximal ratio
transmission (MRT) and zero-forcing (ZF), and adjust the power
of the precoders to compensate for the channel estimation error
uncertainty based on the extreme value theory (EVT) framework.
Finally, we evaluate the performance of our method and show its
superiority concerning worst-case robust precoding, which is used as a benchmark.
  
	\end{abstract}

\section{Introduction}

 Ultra-reliable low-latency communication (URLLC), also known as critical machine-type communications (cMTC), is an essential operation mode in 5G/6G wireless networks \cite{Mahmood.2020}. However, the increasing demand for applications with very strict delay and connectivity requirements makes the network design challenging since achieving reliability and low latency simultaneously is difficult in practice. For instance, factory automation, vehicular communications, and telesurgery may require latency-reliability pairs of (10 ms, $1-10^{-4}$), (1 ms, $1-10^{-5}$) and (1 ms, $1-10^{-9}$), respectively~\cite{popovski2019wireless}. Therefore, efficiently supporting URLLC services requires an accurate statistical model characterization of the operational system, including channel conditions, interference statistics, user mobility, and the behavior of the communication protocols \cite{lopez2022statistical}.
 
 The use of multiple antennas at either one or both sides of a communication system, \textit{e.g.,} single-input multiple-output (SIMO), multiple-input single-output (MISO), and multiple-input multiple-output (MIMO) is a fundamental URLLC enabler \cite{popovski2019wireless}.  Multiple antennas allow performing precoding/combining techniques to improve the signal-to-interference-plus-noise ratio (SINR) by boosting the received signal power, suppressing interference, or even both. This reduces the probability of error when decoding the signal, \textit{i.e.,} higher reliability, and reduces latency since fewer packet re-transmissions are required. However, efficient precoding/combining methods are strictly tied to the availability of channel state information (CSI). Indeed, poor CSI estimations can cause a degradation in the quality of service (QoS) experienced by the user's equipment (UE) since the SINR may fall below the required threshold $\gamma_{th}$. For URLLC, it is particularly important to keep the probability \text{Pr}$\{\text{SINR}<\gamma_{th}\}$ below a stringent permissible error target. Notably, when the reliability requirement is extremely tight, \textit{i.e.,} \text{Pr}$\{\text{SINR}~<~\gamma_{th}\}~\ll1$, classical statistical methods derived from the central limit theorem are not useful as they fail to capture the occurrence of rare error events. Thus, alternative approaches must be considered to overcome this issue, for instance, by exploiting the extreme value theory (EVT) framework.

\subsection{EVT for URLLC}

EVT deals with the stochastic behavior of events that arise in the tail of probability distributions, thus it is a handy tool for URLLC \cite{lopez2022statistical}.   For instance, the authors in \cite{mehrnia2021wireless} presented a methodology to model extreme fade events on the channel. More specifically, they proposed techniques for fitting the tail distribution of the received power to the Generalized Pareto Distribution (GPD), determined the optimal threshold over which the statistics are derived, and calculated the optimal number of samples for fitting the model. The authors in \cite{liu2018ultra} studied a power minimization problem with second-order statistical constraints on latency and reliability. They proposed semi-centralized and distributed queue-aware power allocation techniques for vehicle-to-vehicle communications taking advantage of EVT and Lyapunov stochastic optimization. The work in \cite{mehrnia2022extreme} presented an EVT-based rate selection approach for URLLC. They fitted the tail of the distribution of the received powers to the GPD and determined the maximum transmission rate by including the GPD in the proposed rate selection function. 

\subsection{Related works and Motivation}

In recent years, several works have focused on the solution to minimum-power precoding design problems where the UEs have strict QoS requirements, \textit{e.g.,} SINR, outage probability, data rate \cite{perez2022robust,wang2009worst,zheng2008robust,wang2014outage, medra2016low,medra2016robust,sohrabi2016coordinate,li2022robust}. In this sense, the work in \cite{wang2009worst} proposed a worst-case robust MIMO precoding design to guarantee an SINR performance of the UE for every channel realization. They assumed the channel estimate as the center of an ellipsoid in a multidimensional complex space where the radius is determined by the norm of the CSI error vector, and any channel realization lies inside the ellipsoid. A similar problem was presented in \cite{zheng2008robust} for the multi-user case. The original non-convex problem was re-formulated into semi-definite programming (SDP) form via the S-procedure method and rank relaxations. They also presented an algorithm that extends the robust solution for the multi-user case with both perfect and imperfect CSI at the receiver side while guaranteeing that all the SINRs are above the required target. The authors in \cite{wang2014outage} addressed the minimum-power precoding design problem (hereinafter termed as transmit power minimization problem) with UE's outage constraints. They showed that the probabilistic approach can be converted into a deterministic one with SINR constraints and the same structure as the SDP problem in \cite{zheng2008robust}.
Moreover, the proposed approach allows controlling the radius of the ellipsoid according to the outage demands instead of fixing it to a pre-established value. The work in \cite{medra2016low} also re-arranged the outage constraints into SINR's, specifically for the frequency division duplex (FDD) case. After applying the S-procedure and rank relaxation, the non-convex form was reformulated into a linear objective with linear-matrix inequalities (LMIs) constraints. The work in \cite{medra2016robust} also considered the outage constraints, but with beamforming directions being fixed beforehand. They took advantage of existing precoding methods such as maximal ratio transmission (MRT) or zero-forcing (ZF) and determined the power allocation for each UE. Again, relaxation of the constraints was needed to convert the original problem into an equivalent convex form. The authors in \cite{sohrabi2016coordinate} also considered the transmit power minimization problem with outage constraints by establishing fixed beamforming directions. However, their proposal leads to many outage violations for moderate SINR targets, while the performance was evaluated for outage probabilities higher than $10^{-2}$, which is still far from the most stringent requirements of URLLC.  Finally, the work in \cite{li2022robust} solved the transmit power minimization problem with per-user rate constraints in the finite block length regime. The minimum rates were set to meet specific block error rates in DL transmissions. The original problem was transformed into an SDP problem requiring rank relaxations.

Notice that for the aforementioned minimum-power precoding designs, the authors resorted to approximations or relaxation of constraints that do not fully guarantee to find optimal solutions to the original problem. Some of the approximations are conservative, meaning that the feasible set of precoders of the transformed problem may be smaller than the feasible set of the original formulation. In some cases, the procedures involve LMIs, \textit{e.g.,} \cite{wang2009worst,medra2016low,wang2014outage,sohrabi2016coordinate,li2022robust}, which require high computational and processing costs for the solution. Also, some works, \textit{e.g.,} \cite{sohrabi2016coordinate,wang2014outage,medra2016robust,medra2016low}, evaluate targets that are still far from those required in URLLC applications with strict QoS demands or evaluate the performance with parameters that may not be practical for real applications. The accuracy of the presented approaches for capturing critical events that arise far in the tail of the distributions may be questionable, being EVT a useful tool to overcome this issue.

\subsection{Contributions}
Our work focuses on a minimum-power precoding design to support URLLC in scenarios with imperfect CSI.  Our contributions are four-fold:
\begin{itemize}
    \item We formulate a precoding design optimization problem for transmit power minimization while ensuring URLLC demands at the UEs. We exploit EVT to impose the reliability requirements of the UEs based on the channel estimation and its related uncertainty. Specifically, we fit the data obtained from artificially-generated SINR values to the GPD to model the ultra-reliability region.
    \item We propose an algorithm that leverages state-of-the-art precoding methods to solve the problem. This brings a reduction in the complexity of the problem, thus, reducing the computational costs.
    \item We evaluate the performance of the proposed method using ZF and MRT precoding schemes. We use a worst-case robust precoding scheme as a benchmark to compare the results showing the superiority of our proposed method.
    \item We analyze the impact on the performance of the number of estimation error samples, the confidence when fitting the obtained data to the GPD, the number of URLLC UEs, and the pilot length. We show that there is an optimal pilot length that minimizes the total transmit power, which also increases proportionally with the fitting confidence. Moreover, we show that the fitting confidence must be set larger as the reliability target gets stricter. 
\end{itemize}
   
The work is structured as follows. Section \ref{Sect_System} describes the system model and main assumptions, after which the optimization problem is formulated. In Section \ref{evtO}, we present the EVT-based beamforming design and the proposed algorithm, and discuss MRT and ZF-based implementations. Section \ref{Sect_Bench} presents a benchmark approach to compare with our scheme. In Section \ref{Sect_Numerical}, we illustrate numerical results and validate the proposed algorithm. Finally, Section \ref{section_5} concludes the paper.

\textbf{Notation} Uppercase and lowercase boldface letters denote matrices and vectors, respectively. Superscript $(\cdot)^H$ depicts the Hermitian operator, $(\cdot)^T$ denotes the transpose operator, $(\cdot)^{-1}$ represents the matrix inverse operation, and $||\cdot||$ depicts the norm of a vector. Moreover, $\mathcal{CN}(\mathbf{v},\mathbf{R})$ denotes a complex Gaussian distribution with mean vector $\mathbf{v}$ and covariance matrix $\mathbf{R}$, and $\mathcal{U}(a,b)$ depicts a uniform distribution in the range $[a,b]$. $F_Q(\cdot)$ denotes the cumulative density function (CDF) of the random variable (RV) $Q$
and $\mathbb{i} = \sqrt{-1}$ denotes the imaginary operator. Finally, $\mathcal{Q}(c,D)$ represents the $c\%$-quantile operator of the sample set $D$ and $\mathcal{I}(\cdot)$ denotes the indicator operator. Table \ref{table_0} summarizes the main symbols used throughout the paper.

\begin{table}[t!]
    \centering
    \caption{Main symbols used throughout the paper}
    \label{table_0}
    \begin{tabular}{l  l}
        \hline
        \textbf{Symbol} & \textbf{Definition} \\
            \hline    
            $M$ & number of transmit antennas at the BS\\
            $K$ & total number of UEs\\
            $N$ & number of estimation error vectors available at the BS \\
            $\mathbf{h}_k$ & channel vector between the BS and UE $k$\\
            $\hat{\mathbf{h}}_k$ &  estimate of $\mathbf{h}_k$\\
            
            $\mathbf{e}_k$ & estimation error of $\hat{\mathbf{h}}_k$\\
            
            $\mathbf{s}_k^p$ & pilot sequence transmitted by UE $k$\\
            $\mathbf{s}_k^d$ & data sequence transmitted to UE $k$\\
            $p_{ul}$ & uplink transmit power\\
            $\tau_f$ & length of the data frame\\
            $\tau_e$ & length of the pilot sequence\\
            $\tau_{dl}$ & number of symbols for DL transmission\\   $\mathbf{w}_k$ & precoder intended to UE $k$\\
            $\mathbf{u}_k$ & normalized precoder intended to UE $k$\\
            $\gamma_{k}$ & SINR at UE $k$ \\
            $\gamma_{k}^{tar}$ & SINR target at UE $k$\\

            $\gamma_k^\circ$ &  SINR sample  for UE $k$\\
            $\sigma^2_v$ & noise power\\
        
           $p_k$ & power allocated to UE $k$\\
           $p_{min}$ & minimum transmit power at the BS\\

            $p_{max}$ & maximum transmit power at the BS\\

           $NF$ & noise figure\\

           $BW$ & bandwidth\\
           
            $\Gamma$ & fitting confidence of the GPD\\
            $\rho$ & quantile value of the samples\\
           
          $\zeta_k$ & outage probability target at UE $k$\\

          $\beta_k$ & average channel gain in the link between the BS and UE $k$\\
          $\mathcal{O}_k$& outage probability of UE $k$\\
          $\kappa_k$ & Rician factor of the channel between the BS and UE $k$\\
            \hline
    \end{tabular}
\end{table}

\section{System model}\label{Sect_System}
\begin{figure}[h!]
    \centering
    \includegraphics[width = 0.9\columnwidth]{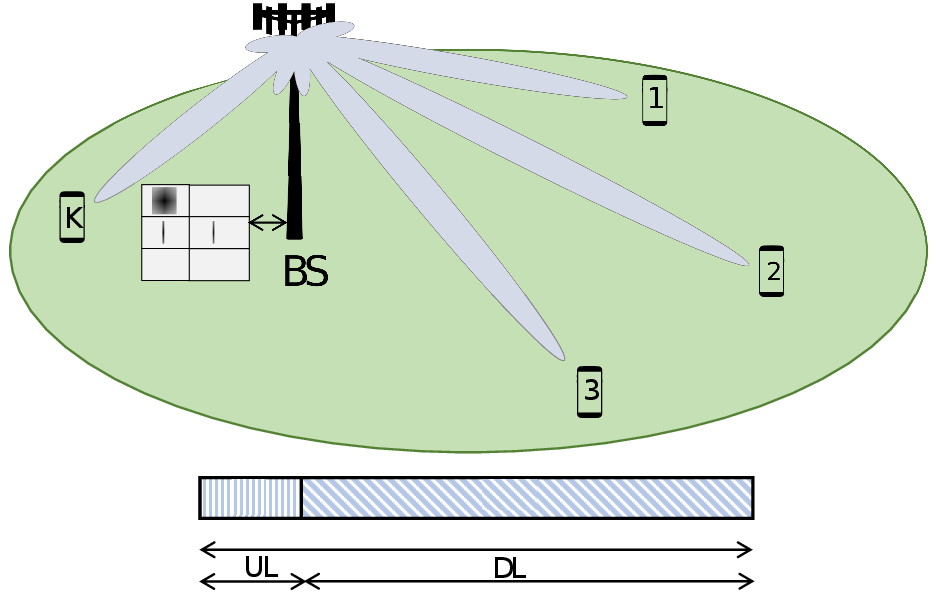}
    \caption{System model and frame structure. The BS at the top of the figure serves a set of $K$ single-antenna URLLC UEs in the DL. $\mathbf{h}_k$ and $\mathbf{w}_k$ represent the channel vector from the BS's antennas to UE $k$ and the precoder vector intended for UE $k$, respectively. The beams from the BS to UE $k$ are formed by precoding the signal through the communication channel, \textit{i.e.,} $\mathbf{h}_k^H\mathbf{w}_k$. Prior to DL transmissions, pilot sequences are transmitted from the UEs to the BS for channel estimation, thus, the BS stores the channel estimates $\hat{\mathbf{h}}_1...\hat{\mathbf{h}}_K$. From previous data, the BS also stores $K$ sets of CSI estimation errors $\mathcal{E}_1...\mathcal{E}_K$. Moreover, the frame structure is displayed at the bottom of the figure with $\tau_f$, $\tau_e$, and $\tau_{dl}$ representing the length of the data frame, length of the pilot sequence, and the number of symbols for DL transmission, respectively.}
    \label{system_model}
\end{figure}

We consider a scenario where a base station (BS) equipped with $M$ antennas serves $K\le M$ single-antenna low-mobility URLLC UEs in the downlink (DL) channel (see Fig. \ref{system_model}). The channels between the BS and the UEs remain constant within a time-frequency coherence block and change independently from block to block. Moreover, channel reciprocity is exploited for channel estimation. Before DL transmissions, the BS estimates the channel coefficients from the $K$ pilots signals of length $\tau_e$ transmitted in the uplink (UL) channel by the UEs. We assume that $\tau_e\ge K$ to guarantee the generation of orthogonal pilots and thus avoid pilot contamination. Let us denote $\tau_{dl}$ as the number of symbols dedicated for DL transmission, therefore $\tau_f=\tau_e+\tau_{dl}$ constitutes the frame duration, while  $\mathbf{h}_k$, $\hat{\mathbf{h}}_k$ $\in$ $\mathbb{C}^{M\times 1}$  are correspondingly the true and estimated channel coefficients between the $k-$th UE and the BS's antennas. 
In practice, estimation errors arise due to noise and uncontrolled interfering signals and cannot be completely removed due to a finite $\tau_e$. We also assume that the BS knows the empirical distribution of the error based on $N$ error samples for any pilot sequence of length $\tau_e$. These samples are utilized together with the estimated channels $\hat{\mathbf{h}}_k$ for precoding design. 
\subsection{Signal model}\label{signalM}

In the UL, the UE $k$ transmits a pilot sequence $\mathbf{s}_k^p\in\mathbb{C}^{\tau_e \times 1}$ with $||\mathbf{s}_k^p||^2 = \tau_e $ such that the signal $\mathbf{Y}\in\mathbb{C}^{M\times \tau_e}$ received at the BS is given by
\begin{align}
\mathbf{Y}=\sum_{k=1}^K\sqrt{p_{ul}}\mathbf{h}_k(\mathbf{s}_k^p)^H+\mathbf{V},
\end{align}
where $p_{ul}$ is the average transmit power of the UEs and $\mathbf{V}\in\mathbb{C}^{M\times \tau_e}$ includes the influence not only of the additive white Gaussian noise (AWGN) but also potential interfering signals at the receiver\footnote{Interference may arise from the use of non-orthogonal (pilot) signals by users in neighboring cells.}. We assume $\mathbb{E}\{\mathbf{V}\}=\mathbf{0}$, which holds in most practical setups as RF signals and AWGN have no direct current level. 
Herein, we adopt the least square (LS) channel estimate
\begin{align}
\hat{\mathbf{h}}_k=\frac{1}{\sqrt{p_{ul}}\tau_e}\mathbf{Y}\mathbf{s}_k^p,
\end{align}
which exploits the fact that the pilot sequences corresponding to different UEs are orthogonal.
Moreover,
\begin{align}\label{error}
    \mathbf{h}_k = \hat{\mathbf{h}}_k + \mathbf{e}_k,
\end{align}
 with $\mathbb{E}\{\mathbf{e}_k\}=\mathbf{0}$ and $\sigma^2_{\hat{h}_k} = \sigma^2_{h_k} - \sigma^2_{e_k}$, where $\mathbf{e}_k$ is the CSI estimation error vector and $\sigma^2_{\hat{h}_k}$, $\sigma^2_{h_k}$, and $\sigma^2_{e_k}$  represent the variances of $\hat{\mathbf{h}}_k$, $\mathbf{h}_k$, and $\mathbf{e}_k$, respectively. Notice that $\sigma_{e_k}^2$ is inversely proportional to the UL SINR and the number of pilot symbols \cite{roy2004maximal}. Also, note that the BS knows $N$ error vectors from previous data, denoted as $\mathcal{E}_k = \{\mathbf{e}_{k,1} \ \mathbf{e}_{k,2}\ ...\ \mathbf{e}_{k,\scaleto{N}{4pt}}\} \ \forall k$.

In the DL, the BS transmits the complex data signal $\mathbf{s}_k^d$ to UE $k$ such that $\mathbb{E}\{||\mathbf{s}_k^d||^2\} = \tau_{dl}$ and $\mathbb{E}\{\mathbf{s}_k^d(\mathbf{s}_i^d)^H\}=\mathbf{0}$ $\forall k\ne i$. Then, the signal $\mathbf{y}_k\in\mathbb{C}^{\tau_{dl}}$ received at UE $k$ is given by
\begin{equation}
   \mathbf{y}_k = \mathbf{h}_k^H\mathbf{w}_k\mathbf{s}_k^d+\sum_{i\neq k}\mathbf{h}_k^H\mathbf{w}_i\mathbf{s}_i^d+\mathbf{v}_k,
\end{equation}
where $\mathbf{w}_k\in \mathbb{C}^{M\times 1}$ depicts the precoding vector intended to UE $k$. Moreover, similar to the UL signal, $\mathbf{v}_k\in\mathbb{C}^{\tau_{dl\times 1}}$, with $\mathbb{E}\{\mathbf{v}_k\}=\mathbf{0}$ and $\mathbb{E}\{||\mathbf{v}_k||^2\}/\tau_{dl}=\sigma_v^2$, comprises the contribution of interference signals and AWGN at the UE $k$. Finally, the SINR at UE $k$ is given by
\begin{equation}\label{eq2}
   \gamma_k(\{\mathbf{w}_k\},\{\mathbf{h}_k\}) = \frac{|\mathbf{h}_k^H\mathbf{w}_k|^2}{\sum_{i\neq k}|\mathbf{h}_k^H\mathbf{w}_i|^2+\sigma_v^2}.
\end{equation}

\subsection{Impact of the estimation error}\label{impactE}
The random fading effect cannot be completely removed due to imperfect CSI estimation. Thus, there is still some remaining randomness in the signal associated with the use of $\hat{\mathbf{h}}_k$. This may prevent the QoS demands from being met as the SINRs may fall below the target $\gamma_k^{tar}$. The impact of the estimation error becomes more severe when the UE transmit power is limited, the average channel gain is low, and/or there is pilot contamination.

\begin{figure}[t!]
    \centering
    \includegraphics[width = 1\columnwidth]{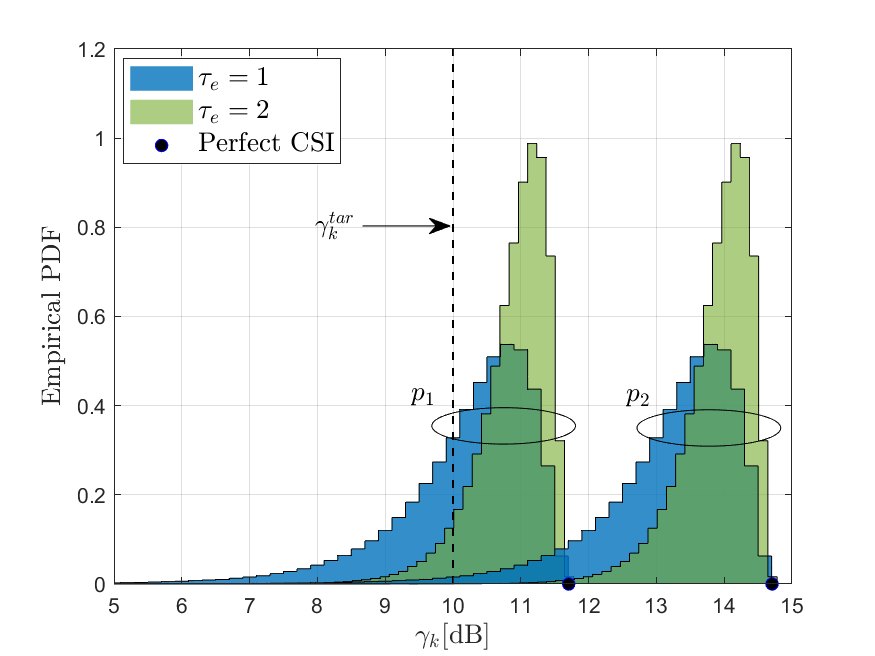}
    
    \caption{Empirical distribution of $\gamma_k$ [dB] for the channel realization $\mathbf{h}_k = \sqrt{10^{-13}}\big[0.118+0.501\mathbb{i}, \ 0.145+0.058\mathbb{i}, \ -0.051+0.022\mathbb{i}, \ 0.087-0.176\mathbb{i}\big]^T$ in Rayleigh fading with  $\tau_e = \{1, 2\}$ and DL transmit power $p_1=23$ dBm and $p_2=26$ dBm. MRT precoding is used with $M = 4$, UL transmit power of 20 dBm, and LS channel estimation.}
    \label{Histograms}
\end{figure}

For example, consider that the BS in the system model serves one UE in the DL, and the minimum SINR to decode the signal with arbitrarily low error probability is $\gamma_1^{tar} = 10$ dB. Fig. 
\ref{Histograms} shows the empirical probability density function (PDF) of the SINR realizations that is achieved for a given channel realization $\mathbf{h}_k$ over $10^6$ channel estimations $\hat{\mathbf{h}}_k$ with $\tau_e = \{1, 2\}$ and LS error estimation. As expected, the variance of the SINR decreases as $\tau_e$ increases. Notably, the probability of falling below the target $\gamma_k^{tar}$ may be high if the precoding, and especially its power allocation ($p_1$), does not consider the estimation error $\big(3.29\times10^{-1}$ and $6.57\times10^{-2}$  for $\tau_e = 1$ and $\tau_e = 2$, respectively, as shown in Fig. \ref{Histograms} (a)$\big)$. Also note that if the transmit power is increased by 3 dB ($p_2$), the SINR realizations are considerably moved to the right and the probability of not meeting the target $\gamma_k^{tar}$ is highly reduced, \textit{e.g.,} $1.83\times 10^{-2}$ and $2.87\times 10^{-4}$ for $\tau_e = 1$ and $\tau_e = 2$, respectively, as depicted in Fig. \ref{Histograms} $(b$). In general, reducing $\gamma_k^{tar}$ and/or increasing the transmit power mitigates the impact of the estimation error on the performance. The reduction of $\gamma_k^{tar}$ decreases the spectral efficiency, leading to a higher transmission latency over the same bandwidth. Therefore, increasing the transmit power seems more appealing if the power budget allows it. However, an arbitrarily high power allocation is not optimal from the energy efficiency point of view and might not guarantee the QoS requirements in the multi-UE case.

\subsection{Problem formulation}
\indent As mentioned earlier, we focus on the precoding design to minimize the transmit power at the BS while ensuring URLLC constraints at each UE $k$. Specifically, we aim to solve the following optimization problem
\begin{subequations}\label{P1}
	\begin{alignat}{2}
	\mathbf{P1:}\qquad &\underset{\{\mathbf{w}_k\}_{\forall k}}{\mathrm{minimize}}  &\ \ \ &\
	\sum_{k=1}^K ||\mathbf{w}_k||_2^2\label{P1:a}\\
	&\text{subject to}   &      & \mathcal{O}_k \leq \zeta_k \ \ \forall k, \label{P1:b}
	\end{alignat}
\end{subequations}

with
\begin{equation}\label{outage_eq}
    \mathcal{O}_k=\text{Pr}\big\{\gamma_k\big(\{\mathbf{w}_k\},\{\mathbf{h}_k\}\big)<\gamma_k^{tar}\big\},
\end{equation}
 where $\gamma_k^{tar}$ depicts the required SINR to achieve a successful transmission and $\zeta_k$ represents the target outage probability at UE $k$. Without loss of generality, we assume $\gamma_k^{tar}=2^{r_k}-1$, where $r_k = B/\tau_{dl}$ and $B$ denotes the number of bits to be transmitted over $\tau_{dl} = \tau_f-\tau_e$ symbols. Notice that the constraint \eqref{P1:b} ensures that the outage probability of UE $k$ is maintained below the target $\zeta_k$. Interestingly, the objective function in \eqref{P1:a} is convex, but we cannot state the convexity of \eqref{P1:b} since the distribution of $\mathbf{h}_k$, and thus the distribution of the SINR, is unknown. Even if the channel distribution is available, the accuracy of the obtained model for capturing events that arise in the tail of the distribution would be low. Therefore, we resort to EVT to reformulate constraint \eqref{P1:b} while proposing a framework that captures rare events and avoids using shape-based models for the channel estimations $\hat{\mathbf{h}}_k$ around the actual channel $\mathbf{h}_k$.

\section{EVT-based
Optimization}\label{evtO}

\subsection{EVT preliminaries}\label{preliminaries}
The main result we exploit from EVT is the following 

\begin{theorem}[Theorem for Exceedances Over Thresholds \cite{coles2001introduction}]
\label{th:exceedance-th-coles}
For an arbitrary RV $X$ from a non-degenerative distribution and for a large enough $\mu$, the cumulative distribution function (CDF) of $Z =X-\mu$ conditioned on $X>\mu$ is given by
\begin{align}\label{GPD}
    F_Z(z) = 1-\bigg[1+\frac{\xi z}{\upsilon}\bigg]^{-\frac{1}{\xi}},
\end{align}
defined on $\{z: z>0 \ \text{and} \ 1+\xi z/\upsilon > 0\}$. The distribution in \eqref{GPD} is known as the \textit{GPD} with shape and scale parameters $\xi$ and  $\upsilon$, respectively.
\end{theorem}

\begin{figure}[t!]
    \centering 
    \includegraphics[width = 1.02\columnwidth]{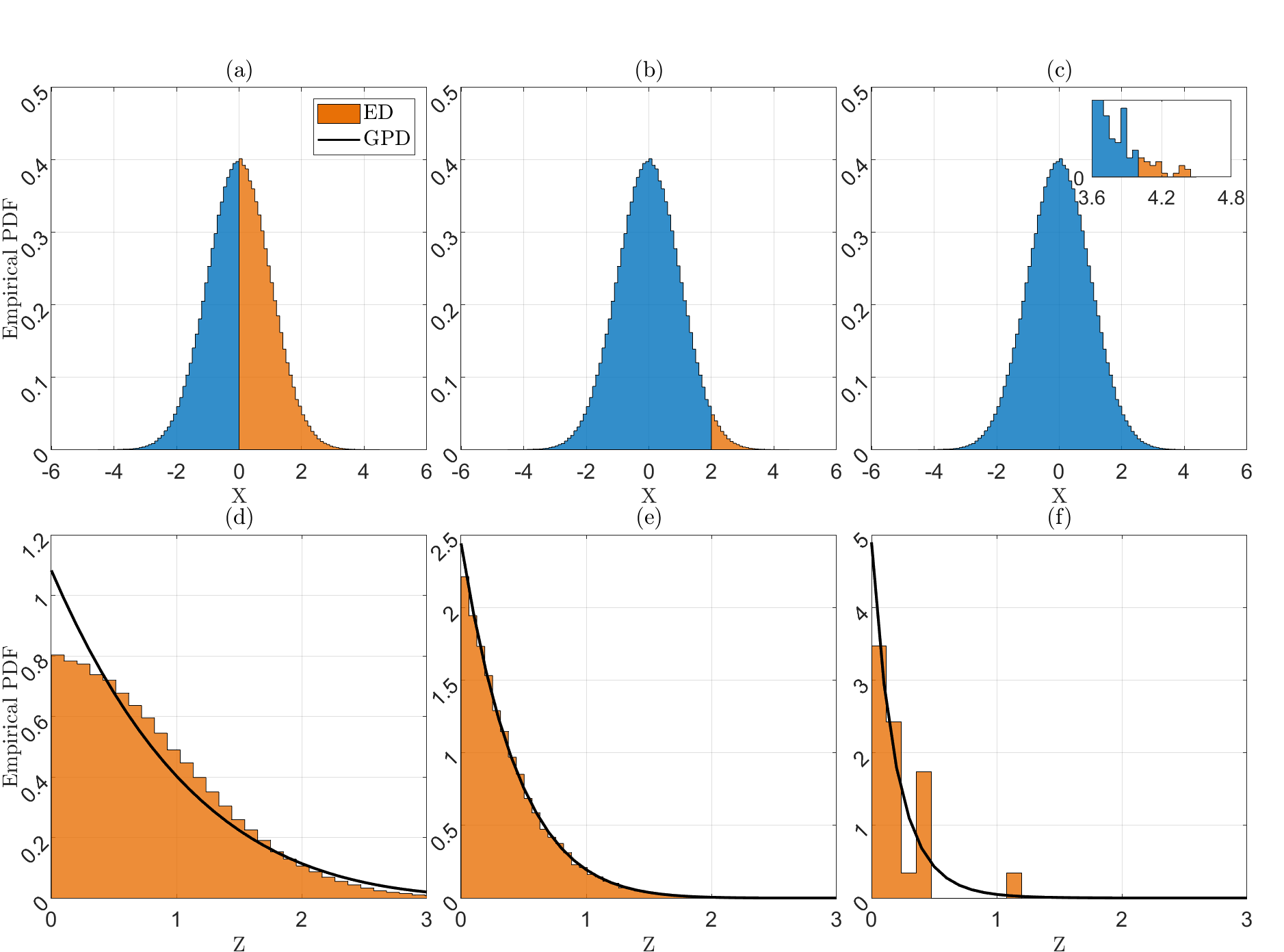}
    \caption{Accuracy of the GPD fitting depending on the selection of $\mu$. The data is drawn from a normal distribution, and the thresholds are $\mu =0, \mu=2$, and $\mu=4$ in the pairs (a)$-$(d), (b)$-$(e), and (c)$-$(f), respectively.}
    \label{GPDfit}
\end{figure}

The parameters of the GPD can be estimated from the available data. Specifically, log-likelihood methods and numerical methods relying on distribution fitting, where the accuracy of the estimates depends on defined confidence levels, are commonly adopted for estimating $\xi$ and $\upsilon$. On the other hand, a \textit{mean residual life plot} may be used to determine a value $\mu_0$, whose \textit{mean residual life function} behaves linearly with respect to $\mu$, and thus, by testing the range of all possible thresholds, it is possible to determine a suitable value for $\mu$ \cite{coles2001introduction}. Another common approach is the so-called \textit{fixed threshold approach} where the threshold is usually set before fitting. In this sense, simple quantile rules have been proposed, \textit{e.g.}, the upper 10\% rule of DuMouchel, which simply uses up to the upper 10$\%$ of the data to fit the GPD, \textit{i.e.,} $\mu = \mathcal{Q}(u,X)$ with $u\ge90\%$  \cite{scarrott2012review,dumouchel1983estimating}. Notice that the selection of $\mu$ is a critical step in the accuracy of the GPD model. On the one hand,  small values of $\mu$ may result in a large number of samples $z$ (large bias), capturing not only events on the tail of the distribution but also values potentially close to the mean, thus affecting the fitting accuracy. On the other hand, large values of $\mu$ may result in a reduced data set (large variance), which would cause an inaccurate parameter estimation. Fig.~\ref{GPDfit} shows the impact of the selection of $\mu$ on the accuracy of the fitting to the GPD fitting. Fig. \ref{GPDfit} (a)$-$(c) show empirical distributions obtained from $10^6$ samples drawn from a normal distribution. The portions of the distribution to the right of the vertical lines (orange color) depict the excess data (ED) over different threshold values. Fig. \ref{GPDfit}  (d)$-$(f) represents the histograms of the ED and the GPD fitting for the thresholds in the corresponding top plots. Notice that for $\mu=0$ and $\mu=4$, the data fitting to the GPD is not accurate. In the former, the ED is large and captures events on the tail and around the mean. At the same time, in the latter, the ED contains only a few samples, leading to an inaccurate fitting/parameter estimation. An accurate fitting is obtained in Fig. \ref{GPDfit} (e) for $\mu = 2$ since the samples are located on the tail and their number is considerably larger than in Fig. \ref{GPDfit} (f).

\subsection{Problem reformulation}
In this subsection, we exploit Theorem~\ref{th:exceedance-th-coles} in Section \ref{preliminaries} and the sets $\mathcal{E}_k \ \forall k$ to rearrange the constraint \eqref{P1:b}. Let us consider an instantaneous channel estimation $\hat{\mathbf{h}}_k$ for UE $k$ and the channel estimation error history for the corresponding channel $\mathcal{E}_k$. Because of the zero-mean properties of $\mathbf{e}_k$, the distributions of $\mathbf{h}_k$ and $\hat{\mathbf{h}}_k$ share the same mean but differ in their variances. Thus, real and imaginary components of $\mathbf{h}_k$ lie around the mean of real and imaginary components of $\hat{\mathbf{h}}_k$. Therefore, by adding up each entry of the error set $\mathcal{E}_k$ to $\hat{\mathbf{h}}_k$, we obtain the new set
\begin{align}\label{set_H}
    \mathcal{H}_k = \mathcal{E}_k + \hat{\mathbf{h}}_k.
\end{align}
Notice that this is possible due to the independence between the error $\mathbf{e}_k$ and channel estimate $\hat{\mathbf{h}}_k$ enabled by the LS estimation method. Moreover, $\mathcal{H}_k = \{\Tilde{\mathbf{h}}_{k,1}, \Tilde{\mathbf{h}}_{k,2},...,\Tilde{\mathbf{h}}_{k,\scaleto{N}{4pt}}\}$, with $\Tilde{\mathbf{h}}_{k,n}=\mathbf{e}_{k,n}+\hat{\mathbf{h}}_k$, is a group of possible channel realizations for the link between UE $k$ and the BS that may have led to a channel estimation $\hat{\mathbf{h}}_k$. This implies that the larger the value of $N$, the smaller (probabilistically) the difference between at least one element $\Tilde{\mathbf{h}}_{k, n}$ in $\mathcal{H}_k$ and the actual $\mathbf{h}_k$. Additionally, the set $\mathcal{H}_k$ also contains entries that are farther from $\mathbf{h}_k$, which are beneficial for mimicking the poorest estimation cases (left tails in Fig. \ref{Histograms}). 

For the $n-$th entry $\Tilde{\mathbf{h}}_{k,n}$ of the set $\mathcal{H}_k$, we can generate a  sample of the SINR of UE $k$ using \eqref{eq2} as follows 
\begin{equation}\label{SNR-H-estimates}
   \gamma^\circ_{k,n}(\!\{\mathbf{w}_k\},\!\{ \Tilde{\mathbf{h}}_{k,n}\}\!) \!=\! \frac{|\Tilde{\mathbf{h}}_{k,n}^H\mathbf{w}_k|^2}{\sum_{i\neq k}|\Tilde{\mathbf{h}}_{k,n}^H\mathbf{w}_i|^2+\sigma_v^2}. 
\end{equation}
 To meet the reliability demands, we must ensure that for a channel estimation $\hat{\mathbf{h}}_k$, most of the samples satisfy $ \gamma^\circ_{k,n}(\{\mathbf{w}_k\},\{ \Tilde{\mathbf{h}}_{k,n}\})>\gamma_k^{tar}$. However, in this case, the data of interest is located on the left tail of the distribution (samples that do not meet the SINR requirements). Still, to apply EVT, specifically the Theorem~\ref{th:exceedance-th-coles}, we must have the data on the right tail. Thus, we may proceed with a simple transformation of \eqref{P1:b} as follows 
\begin{align}\label{step1}
    \text{Pr}\big\{\gamma^\circ_{k,n}\big(\{\mathbf{w}_k\},&\{\Tilde{\mathbf{h}}_{k,n}\}\big)<\gamma_k^{tar}\big\} 
    \nonumber\\
    &=\text{Pr}\bigg\{\frac{1}{\gamma^\circ_{k,n}\big(\{\mathbf{w}_k\},\{\Tilde{\mathbf{h}}_{k,n}\}\big)}>\frac{1}{\gamma_k^{tar}}\bigg\}.
\end{align}
 Some samples of the RV $1/\gamma^\circ_{k,n}\big(\{\mathbf{w}_k\},\{\Tilde{\mathbf{h}}_{k,n}\}\big)$ might be significantly dispersed from the rest which may affect the fitting to the ED that we will perform in the next steps. To mitigate this issue, we introduce a concave transformation $f(\cdot)$ to all samples as
\begin{equation}\label{step2}    
\text{Pr}\bigg\{f\bigg(\frac{1}{\gamma^\circ_{k,n}\big(\{\mathbf{w}_k\},\{\Tilde{\mathbf{h}}_{k,n}\}\big)}\bigg)>f\bigg(\frac{1}{\gamma_k^{tar}}\bigg)\bigg\}.
\end{equation} 
Let us now define
\begin{align} 
\psi_k\big(\{\mathbf{w}_k\},\{\Tilde{\mathbf{h}}_{k,n}\}\big)&\triangleq f\big(1/\gamma^\circ_{k,n}(\{\mathbf{w}_k\},\{\Tilde{\mathbf{h}}_{k,n}\})\big),\label{InvFunct}\\
\phi_k &\triangleq f\big(1/\gamma_k^{tar}\big),\label{Invtarget}
\end{align}
for ease of notation. We can now set a threshold $\mu_k$ and apply the definition of  conditional probability as 
\begin{align}\label{step3}
     \text{Pr}&\Big\{\psi_k\big(\{\mathbf{w}_k\},\{\Tilde{\mathbf{h}}_{k,n}\}\big)>\phi_k\Big
    \}\nonumber\\
    &=
     \text{Pr}\Big\{\psi_k\big(\{\mathbf{w}_k\},\!\{\Tilde{\mathbf{h}}_{k,n}\}\big)\!>\!\mu_k\Big\}\text{Pr}\Big\{\!\psi_k\big(\!\{\mathbf{w}_k\},\!\{\Tilde{\mathbf{h}}_{k,n}\}\!\big)\!-\!\mu_k\nonumber\\
     &\qquad\qquad\ \ >\phi_k-\mu_k \big| \psi_k\big(\{\mathbf{w}_k\},\{\Tilde{\mathbf{h}}_{k,n}\}\big)>\mu_k\Big\}.
\end{align}
Moreover, according to DuMouchel's rule, we can set the threshold $\mu_k=\mathcal{Q}(\rho\times 100, \psi_k)$, thus as a function of $\{\mathbf{w}_k\}$ and $\{\Tilde{\mathbf{h}}_{k,n}\}$ such that 
\begin{align}\label{step4}
\text{Pr}\Big\{\psi_k &\big(\{\mathbf{w}_k\},\{\Tilde{\mathbf{h}}_{k,n}\}\big)\!>\!\mu_k\big(\{\mathbf{w}_k\},\{\Tilde{\mathbf{h}}_{k,n}\}\big)\Big\}\nonumber\\
&\approx
\!\frac{1}{N}\sum_{n=1}^N\mathcal{I}\Big[\psi_k\big(\{\mathbf{w}_k\},\{\Tilde{\mathbf{h}}_{k,n}\}\big)\!>\!\mu_k\big(\{\mathbf{w}_k\},\{\Tilde{\mathbf{h}}_{k,n}\}\big)\Big]\nonumber\\
&=
 1-\rho
\end{align}
 holds.
Thus, we have that
\begin{align}\label{step5}
\text{Pr}\Big\{\gamma_k &\big(\{\mathbf{w}_k\}\big)<\gamma_k^{tar}\Big\} \nonumber\\
&= (1\!-\!\rho)\Big(1\!-\!F_{Q_k}\big(\phi_k\!-\!\mu_k(\{\mathbf{w}_k\},\{\Tilde{\mathbf{h}}_{k,n}\})\big)\Big),
\end{align}
where 

\begin{align}\label{GPDdata}
    Q_k\triangleq\Big(\psi_k &\big(\{\mathbf{w}_k\},\{\Tilde{\mathbf{h}}_{k,n}\}\big)-\mu_k\big(\{\mathbf{w}_k\},\{\Tilde{\mathbf{h}}_{k,n}\}\big)\ \nonumber\\
    &\Big| \psi_k\big(\{\mathbf{w}_k\},\{\Tilde{\mathbf{h}}_{k,n}\}\big)>\mu_k\big(\{\mathbf{w}_k\},\{\Tilde{\mathbf{h}}_{k,n}\}\big)\Big).
\end{align}
Next, we proceed to fit all data samples $Q_k$
to the GPD in \eqref{GPD} to obtain the estimates $\hat{\upsilon}_k(\{\mathbf{w}_k\},\{\Tilde{\mathbf{h}}_{k,n}\}) = \hat{\upsilon}_k$ and $\hat{\xi}_k(\{\mathbf{w}_k\},\{\Tilde{\mathbf{h}}_{k,n}\}) = \hat{\xi_k}$ of the parameters $\upsilon$ and $\xi$  with $z = \phi_k -\mu_k\big(\{\mathbf{w}_k\},\{\Tilde{\mathbf{h}}_{k,n}\}\big)$. With the estimates, \eqref{outage_eq} can be re-written as
\begin{equation}\label{Outage_eq_2}
    \mathcal{O}_k = (1-\rho)\Big(1+\frac{\hat{\xi}_k}{\hat{\upsilon}_k}\big(\phi_k -\mu_k(\{\mathbf{w}_k\},\{\Tilde{\mathbf{h}}_{k,n}\})\big)\Big)^{-1/\hat{\xi}_k}.
\end{equation}
Notice that after the transformation of $\mathcal{O}_k$, \textbf{P1} remains as a non-convex problem which is yet challenging to solve. We address this issue in the next subsection.
\subsection{Proposed algorithm}
 Note that common non-convex optimization solvers such as those based on genetic and particle swarm algorithms might not often provide feasible solutions to \textbf{P1} because of the high non-linearity of the constraints, the difficulty to properly configure the optimization hyperparameters, and the large amount of required computational resources. Similar to prior work, \textit{e.g.,} \cite{perez2022robust, medra2016robust,sohrabi2016coordinate}, that fixed the precoding directions for reducing complexity, we propose an algorithm that exploits state-of-the-art linear precoding schemes, \textit{e.g.,} ZF and MRT, for transmit power minimization. 
 
 First, we depart from the channel estimations $\hat{\mathbf{h}}_k$ to compute the precoders as $\mathbf{w}_k=\sqrt{p_k}\mathbf{u}_k$ with $p_k$ as the power allocated to UE $k$ which is initially set to a minimum value $p_{min}$ to all UEs. The normalized precoding directions are given by
 \begin{align}\label{precoding}
     \mathbf{u}_k & =\frac{\mathbf{z}_k^*}{||\mathbf{z}_k||},
 \end{align}
 where $\mathbf{z}_k = \hat{\mathbf{h}}_k$ for MRT precoding, and $[\mathbf{z}_1,   \mathbf{z}_2,...,\mathbf{z}_k]=\hat{\mathbf{H}}(\hat{\mathbf{H}}^H\hat{\mathbf{H}})^{-1}$ for ZF precoding with $\hat{\mathbf{H}}= [\hat{\mathbf{h}}_1,\hat{\mathbf{h}}_2,...,\hat{\mathbf{h}}_k]$. 
 Then, compute the sets $\mathcal{H}_k$ according \eqref{set_H}, and for every UE $k$, compute \eqref{SNR-H-estimates}, \eqref{InvFunct} and \eqref{Invtarget}. Next, determine the value of $\mu_k$ as the $\rho-$quantile ($\%$) of $\psi_k$ such that \eqref{step4} holds. Then, compute the excesses $Q_k$ in \eqref{GPDdata} and estimate the parameters of the GPD with confidence $\Gamma$ via log-likelihood estimation to obtain bounds as $[\hat{\upsilon}_{k,\scaleto{LB}{4pt}},\hat{\xi}_{k,\scaleto{LB}{4pt}}]$ and $[\hat{\upsilon}_{k,\scaleto{UB}{4pt}},\hat{\xi}_{k,\scaleto{UB}{4pt}}]$. With the upper estimate, proceed to obtain an upper outage probability bound $\mathcal{O}_{k, UB}$ by evaluating the pair  $[\hat{\upsilon}_{k,\scaleto{UB}{4pt}},\hat{\xi}_{k,\scaleto{UB}{4pt}}]$ in \eqref{Outage_eq_2}. Then, if the bound is above the target $\zeta_k$, the power $p_k$ is increased in a small value $\Delta p$, and the process starts again from the computation in \eqref{SNR-H-estimates}. Nevertheless, if the outage 
 bound is below the target $\zeta_k$, the real outage probability will also be below $\zeta_k$ if the parameters $\rho$ and $\Gamma$ are properly configured. In such a case, a similar analysis must be done with the remaining UEs until the outage bounds for all UEs are below their respective targets $\zeta_k$ simultaneously or until the total power constraint $p_{max}$ is violated, and there is no feasible solution. Notice that the selection of $\Delta p$ significantly impacts the performance of the proposed algorithm. On the one hand, large values may cause the algorithms not to find solutions to the problem, while small values will make the processing time larger. Nevertheless, it is recommended to select a small value that ensures finding the solutions, \textit{e.g.}, $[-35,-15]$ dBm. Finally,  the precoders $\mathbf{w}_k = \sqrt{p_k}\mathbf{u}_k \ \forall k$ constitute the solution to \textbf{P1}. 

\textbf{Algorithm 1} summarizes the previously discussed steps. The algorithm also comprises the initialization of the transmit powers $p_k$ to a minimum power $p_{min}$ and the upper bounds in the outage probability $\mathcal{O}_{k,\scaleto{UB}{4pt}} =1$ in lines 1 and 4, respectively. The value of $p_{min}$ is recommended to be small, \textit{e.g.,} $-$30 dBm, and $p_{max}$ can be chosen according to hardware constraints, \textit{e.g.,} $46-47$ dBm, in typical BSs \cite{ahmadi20195g}.
\begin{algorithm}[t!]
\caption{Robust Minimum-Power Precoding for URLLC.}
\hspace*{\algorithmicindent} \textbf{Inputs:} $\rho,\ \Gamma, \  \{\hat{\textbf{h}}_{k}\},\ \{\mathcal{E}_k\}$, $p_{min}$, $p_{max}, \Delta p$  \\
\hspace*{\algorithmicindent}
\textbf{Outputs:} $\ \{\textbf{w}_k\}$
\begin{algorithmic}[1]
\State Initialize $p_k\gets p_{min} \ \forall k$
\State Compute $\mathcal{H}_k 
\ \forall k$ according to \eqref{set_H}
\State Compute $\mathbf{u}_k$ and $\phi_k \ \forall k$ according to \eqref{precoding} and \eqref{Invtarget}, respectively 
\State Initialize the outage bound  $\mathcal{O}_{k,\scaleto{UB}{4pt}} = 1\ \forall k$
\While{$\mathcal{O}_{k,\scaleto{UB}{4pt}}  > \zeta_k \ \forall k \ \textbf{and} \ \sum_{k =1}^K p_k \le p_{max}$}
\State For UE $k$ compute $\gamma_{k,\scaleto{N}{4pt}}$ and $\psi_k$ according to \eqref{SNR-H-estimates} and \eqref{InvFunct}, respectively
\State Find $\mu_k\big(\{\mathbf{w}_k\},\{\Tilde{\mathbf{h}}_{k,n}\}\big)$ as the $\rho-$quantile of $\psi_k$
\State Compute the data $Q_k$ in \eqref{GPDdata}
\State Fit the GPD to $Q_k$ with   confidence $\Gamma$ to obtain $[\hat{\upsilon}_{k,\scaleto{LB}{4pt}},\hat{\xi}_{k,\scaleto{LB}{4pt}}],[\hat{\upsilon}_{k,\scaleto{UB}{4pt}},\hat{\xi}_{k,\scaleto{UB}{4pt}}]$
\State Evaluate $[\hat{\upsilon}_{k,\scaleto{UB}{4pt}},\hat{\xi}_{k,\scaleto{UB}{4pt}}]$ in \eqref{Outage_eq_2} to obtain $\mathcal{O}_{k,\scaleto{UB}{4pt}}$

\If{$\mathcal{O}_{k,\scaleto{UB}{4pt}} < \zeta_k$} 
\State $\mathbf{w}_k = \sqrt{p_k}\mathbf{u}_k$
\State pick another UE $k$
\Else
\State $p_k = p_k + \Delta p$
\EndIf
\EndWhile
\end{algorithmic}
\end{algorithm}
\section{Benchmark approach}\label{Sect_Bench}
As a benchmark, we consider the work in \cite{wang2009worst}, where the authors solved the transmit power minimization problem with SINR constraint for a single UE MIMO system given by
\begin{subequations}\label{P3}
	\begin{alignat}{2}
	\mathbf{P2 a:} \ &\underset{\mathbf{W}_k	\succeq 0}{\mathrm{minimize}}  & \ &\
	\text{Tr}(\mathbf{W}_k)\label{P3:a}\\
	&\text{subject to}   &      & \gamma_k(\mathbf{W}_k,\hat{\mathbf{h}}_k+\mathbf{e}_k)\!<\!\gamma_k^{tar} \ \forall  \mathbf{e}_k\!:\! ||\mathbf{e}_k||\le \epsilon, \label{P3:b}
	\end{alignat}
\end{subequations}
where $\mathbf{W}_k = \mathbf{w}_k\mathbf{w}_k^H$ and $||\mathbf{e}_k||\le \epsilon$ ensures that all possible channels $\mathbf{h}_k$ in $\mathbb{C}^{M}$ lie inside an ellipsoid centered at the estimated channel $\hat{\mathbf{h}}_k$ with radius $\epsilon$. To guarantee a certain level of reliability $\zeta_k$, it is enough to control the radius $\epsilon$ of the ellipsoid such that $100\times (1-\zeta_k)\%$ of the channels $\mathbf{h}_k$ lie inside the boundaries. This can be achieved by defining $\epsilon=\mathcal{Q}\big(100\times(1-\zeta_k), ||\mathcal{E}_k||\big)$ which also imposes a minimum number of required samples $N=1/\zeta_k$ in the set $\mathcal{E}_k$ to effectively determine the quantile. \textbf{P2a} is not convex in its current form, therefore, it is transformed into the equivalent convex SDP problem \cite{wang2009worst}
\begin{subequations}\label{P4}
	\begin{alignat}{2}
	\mathbf{P2b:}\ &\underset{\mathbf{W}_k, \mathbf{Z}, \Omega}{\mathrm{minimize}}  &\  &\
	\text{Tr}(\mathbf{W}_k)\label{P4:a}\\
	&\text{subject to}   &      & \text{Tr}\big[(\textbf{Z}\!-\!\textbf{W}_k)\hat{\mathbf{h}}_k^H\hat{\mathbf{h}}_k\big]\!+\!\epsilon^2\Omega\!+\!\gamma_k^{tar}\!\le 0\label{P4:b},\\
&  &  & \begin{bmatrix}
\mathbf{Z} & \mathbf{W}_k\\
\mathbf{W}_k & \mathbf{W}_k + \Omega\mathbf{T} 
\end{bmatrix}	\succeq 0 \label{P4:c},\\
& & & \mathbf{W}_k	\succeq 0, \label{P4:d}\\
& & &  \Omega\ge 0, \label{P4:e}
	\end{alignat}
\end{subequations}
where $\mathbf{Z}$ and $\Omega$ depict auxiliary variables, and $\mathbf{T} = \Upsilon\mathbf{I}$ with $\mathbf{I}$ as the identity matrix. $\Upsilon$ determines the shape of the ellipsoid being a sphere for the case $\Upsilon = 1$. The complexity of this problem grows with the number of variables $2M^2+1$ in polynomial time \cite{monteiro2000polynomial}, and the solution can be found using common solvers/algorithms such as CVX or Interior point methods (IPM).

 The main disadvantages of this approach are related to the use of LMIs and the computation complexity for finding the solution. Moreover, the need for at least $1/\zeta_k$ samples to efficiently compute $\epsilon$ represents another key drawback.

\section{Numerical results}\label{Sect_Numerical}
 
In this section, we evaluate the performance of the proposed algorithm for a single URLLC UE and multiple URLLC UEs. We consider that the BS is equipped with a uniform linear array and assume the spatially-correlated Rayleigh fading model for the channels where $\mathbf{h}_k\sim \mathcal{CN}(\mathbf{0},\mathbf{R}_k)$. $\beta_k = \frac{1}{M}\text{tr}(\mathbf{R}_k)$  denotes the average channel gain accounting only for path loss and $\mathbf{R}_k$ represents the spatial correlation matrix. For $\mathbf{R}_k$, we adopt the local scattering spatial correlation model and its approximation for Gaussian angular distribution with half-wavelength antenna separation~\cite{bjornson2017massive} 
\begin{align}\label{R_model}
   [\mathbf{R}_k]_{t,m} = \frac{\beta_k}{L}\sum_{l=1}^L &\exp({\pi \mathbb{i}(t-m)\sin{\varphi_{k,l}}})\nonumber\\
   \times&\exp({-\frac{1}{2}\sigma_{\varphi_k}^2\pi(t-m)\cos{\varphi_{k,l}}}),
\end{align}
 where $L$ denotes the number of multi-path clusters, $\varphi_{k,l} \sim \mathcal{U}\big(\bar{\varphi}_k-\frac{2\pi}{9},\ \bar{\varphi}_k+\frac{2\pi}{9}\big)$ is the nominal angle of arrival of cluster $l$ for UE $k$, which is uniformly distributed around the azimuth angle of the UEs relative to the bore-sight of the BS antenna array ${\bar{\varphi}}_k$. Moreover, $\sigma_{\varphi_k}$
 depicts the angular standard deviation of the paths within a multi-path cluster. The estimation error is assumed to be distributed as $\mathbf{e}_k\sim\mathcal{CN}(\mathbf{0},\frac{\sigma^2_n}{p_{ul}\tau_e}\mathbf{I})$, which corresponds to the scenario without pilot contamination \cite{roy2004maximal}. All UEs are assumed to have the same average channel gain $\beta_k$ and UL transmit power $p_{ul}$ for simplicity in the modeling, but  $\bar{\varphi}_k\sim\mathcal{U}(0,2\pi)$. The noise power is given by $\sigma_n^2 = -173.8+10\log_{10}BW+ NF$ dBm where $BW$ and $NF$ represent the communication bandwidth and noise figure, respectively.  Moreover, in the EVT-based approach, we use $f = 10\log_{10}(\cdot)$  as the concave function. All simulation parameters are displayed in Table \ref{table_2} and are based on \cite{wang2009worst,bana2018ultra,dumouchel1983estimating,bjornson2017massive,ahmadi20195g}. Finally, \textbf{P2b} for the benchmark approach is solved using CVX tool.

\begin{figure}[h!]
    \centering
    \includegraphics[width = 1\columnwidth]{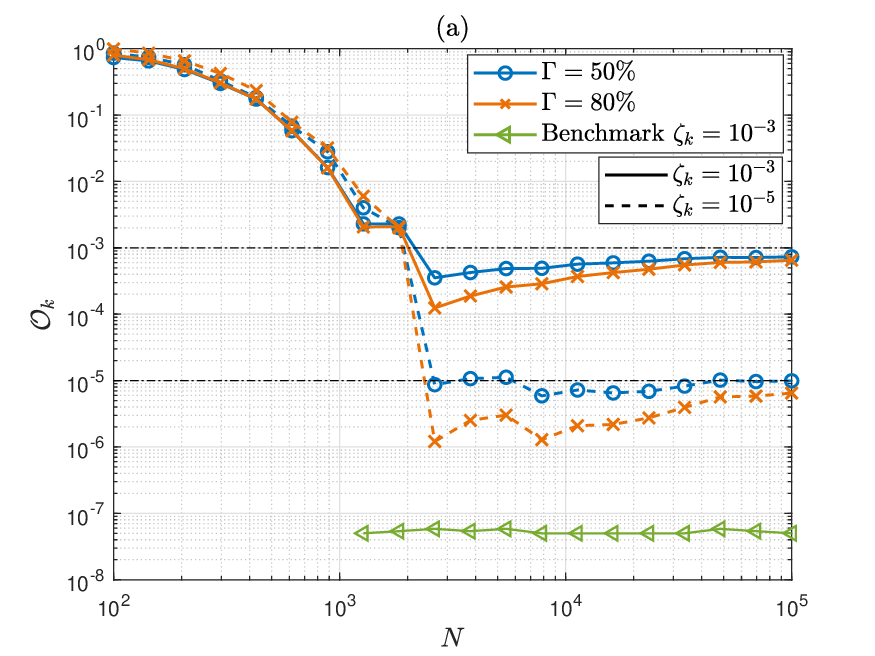}
    \includegraphics[width = 1\columnwidth]{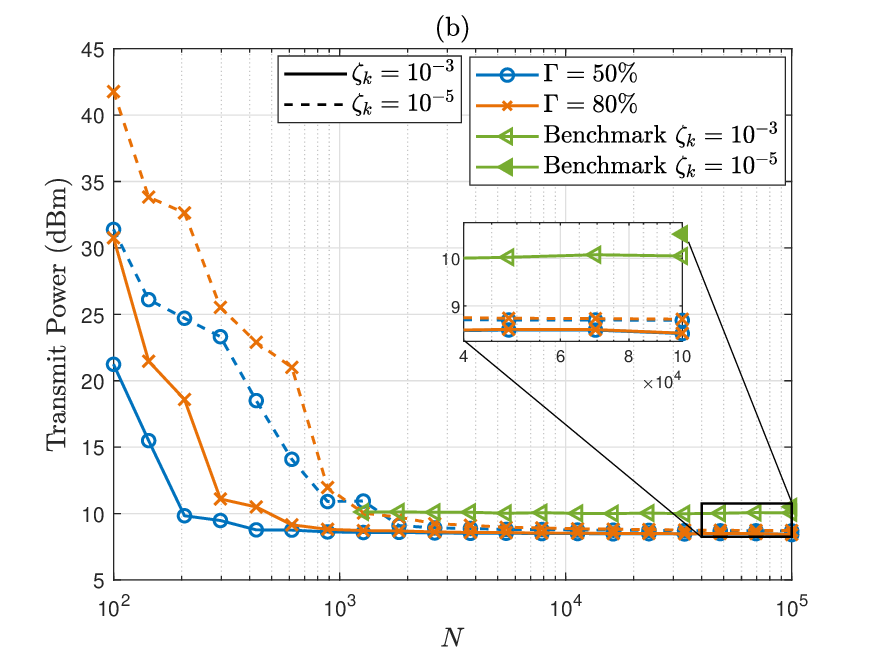}
    \caption{Outage probability (a) and transmit power (b) probability as a function of the number of estimation error samples. We set $M=8$ and $\tau_e=1$, and employ MRT for the EVT approach.}
    \label{fig:2}
\end{figure}
\subsection{Performance evaluation for single URLLC UE}
Fig. \ref{fig:2} (a) shows the achievable outage probability as a function of the number of error vectors  $N$ in a single UE scenario with a pilot length $\tau_e = 1$. Notice that the outage probabilities attained by our approach are far above the outage target $\zeta_k$ when exploiting only a small number of samples, approximately $N\le2000$. This is because the algorithm cannot always find feasible solutions when the length of the sets $Q_k$ is small due to inaccurate GPD fittings; thus, in those events, we declare an outage. On the other hand, for larger $N$ and high-reliability targets, \textit{e.g.,} $10^{-5}$, the actual outage probabilities are below $\zeta_k$ especially if the fitting confidence is sufficiently high, \textit{e.g.,} $\Gamma= 80 \%$. The benchmark scheme meets the outage requirements amply for both targets when $N\ge1/\zeta_k$ since that represents the minimum number of required samples for finding the quantiles as discussed in Section \ref{Sect_Bench}.  In fact, the outage probability values in the case of $\zeta_k = 10^{-5}$ are not displayed as they are smaller than $10^{-8}$ and thus difficult to estimate due to computational resource limitations.  Notice that the benchmark approach achieves lower outage probabilities compared to our approach but at the cost of higher power consumption as we discuss next.   Fig. \ref{fig:2} (b) focuses the analysis on required transmit power as a function of $N$. We can observe regions of instability and high power consumption given a relatively small $N$ for the EVT-based scheme. In contrast, the transmit power converges as $N$ increases, experiencing a small reduction as $N\xrightarrow{}\infty$. The power requirements of the benchmark approach increase slightly with $N$ and are more than 1 dB above our approach's requirements in the stability region. Notice that for $\zeta_k= 10^{-5}$ there is only one feasible point at $N = 10^5$ since that is the minimum number to compute the quantile. The figure shows that our method's main advantage is reducing the transmit power while taking the outage probabilities as close as possible to the targets, which is not achieved by the benchmark approach. We assume  $N = 10^4$ for the remaining simulations.
\begin{table}[t!]
    \centering
    \caption{Simulation parameters}
    \label{table_2}
    \begin{tabular}{c c}
        \hline
        \textbf{Parameter} & \textbf{Value} \\
            \hline           
            $M$ & 4, 8 \\
            $K$ &  $1-5$\\
            $p_{min}$ & $-30$ dBm \\ $p_{max}$ &  47 dBm \cite{ahmadi20195g}\\
            $N$  & $10^4$ \\
            $NF$ &  7 dB \cite{bjornson2017massive}\\
            $BW$ & $60$ kHz \cite{bana2018ultra} \\ $\tau_f$ &  42 (OFDM symbols)$-0.75$ ms \cite{bana2018ultra} \\
            $\tau_e$ & $1-8$ \cite{bana2018ultra} \\
            $B$ & 256 \cite{bana2018ultra}\\
            $\rho$ & $0.95$ \cite{dumouchel1983estimating}\\ 
            $\Gamma$ &  $10-90\%$\\
            $\beta_k$ & $-115$ dB \cite{bjornson2017massive} \\ $\zeta_k$ &  $10^{-6}-10^{-1}$ \\
            $\sigma_{\varphi_k}$ & $\frac{\pi}{36}$ \cite{bjornson2017massive} \\ $L$ & 10 \\ 
            $p_{ul}$ & $20$ dBm \\ $\Upsilon$& 1 \cite{wang2009worst} \\
             
            \hline
    \end{tabular}
\end{table}

\begin{figure}[h!]
    \centering
    \includegraphics[width = 1\columnwidth]{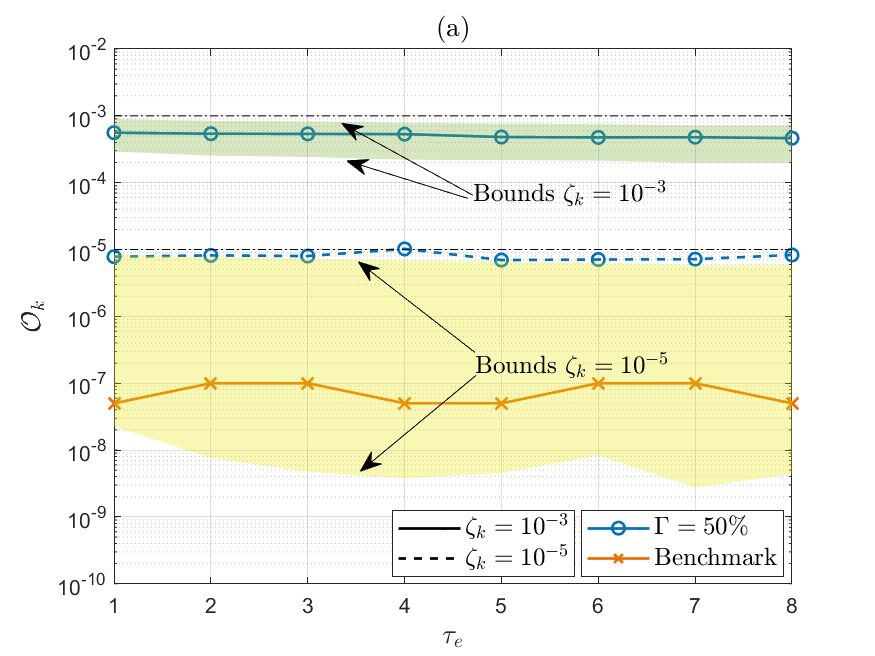}
    \includegraphics[width = 1\columnwidth] {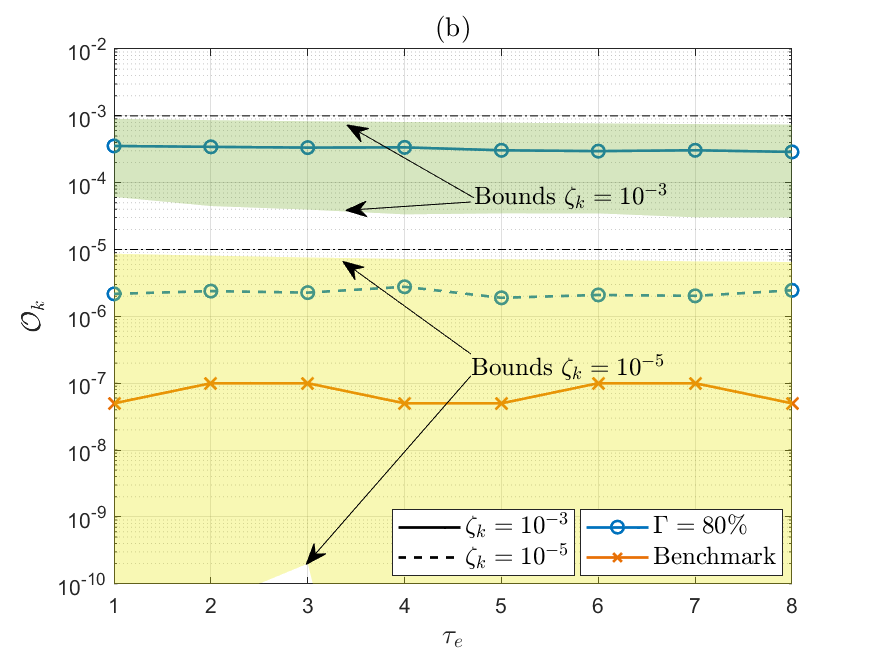}
    \caption{Outage probability and confidence bounds for $\Gamma = 50\%$ (a) and $\Gamma = 80\% $ (b)  as a function of the pilot lengths  $\tau_e$. We set $M=8$ and $N=10^4$, and employ MRT for the EVT approach. Green and yellow regions represent the outage probabilities between lower and upper bounds $\mathcal{O}_{k,LB}$ and $\mathcal{O}_{k,UB}$, respectively.}
    \label{fig:3}
\end{figure}

Figs. \ref{fig:3} (a) and (b) show the outage probability and bound regions versus $\tau_e$  for $\Gamma = 50\%$ and $\Gamma = 80\%$, respectively. The figures also show the outage probabilities achieved by the benchmark approach. Notice that the bound regions get wider as the fitting confidence increases and that for low fitting confidence, \textit{e.g.,} $\Gamma\le 50 \%$, the outage constraint is more likely to be violated, especially for smaller $\zeta_k$ as it is the case of $\zeta_k=10^{-5}$ in Fig. \ref{fig:3} (a). Thus, a higher $\Gamma$ may be required to meet the reliability requirements in practice.

\begin{figure}[t!]
    \centering
    \includegraphics[width = 1\columnwidth]{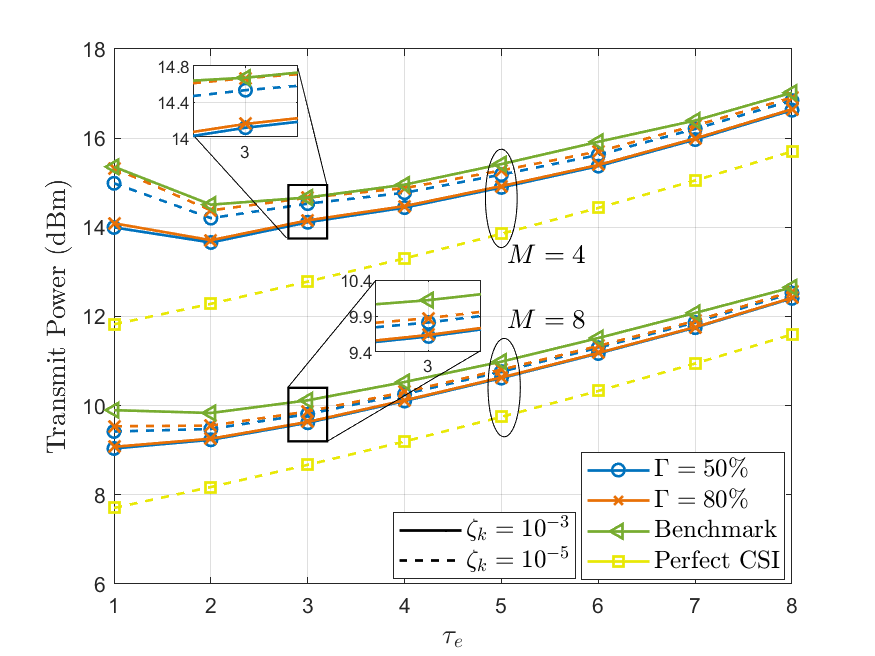}

    \caption{Transmit power as a function of the pilot lengths $\tau_e$. We set $M\in~\{4,8\}$ and $N=10^4$, and employ MRT for the EVT approach.}
    \label{fig:4}
\end{figure}

Fig. \ref{fig:4} shows the transmit power required for achieving $\zeta_k=10^{-3}$ and $\zeta_k=10^{-5}$ as a function of the pilot lengths given $M\in\{4,8\}$ and $\Gamma\in\{50\%,80\%\}$. Notice that the gap between the transmit power for $\Gamma=50\%$ and $\Gamma=80\%$ increases with the reliability level, being larger for $\zeta_k = 10^{-5}$. Interestingly, there is a pilot length that minimizes the transmit power depending on the number of antennas $M$. This is because the estimation error may be significant given a relatively small $\tau_e$, leading to higher power requirements to achieve a certain SINR. On the other hand, a relatively large $\tau_e$ implies better channel estimation and a smaller $\tau_{dl}$, consequently higher SINR  requirements and thus transmit power. For the specific results illustrated in Fig.~\ref{fig:4}, when the number of antennas is $M = 4$, the diversity and degrees of freedom (DoF) gains of the system are low, which implies that a better channel estimation is required to achieve the requirements, \textit{i.e.,} $\tau_e>1$. For $M =8$, the system takes advantage of extra diversity and DoF gains offered by the additional four antennas, thus, optimally meeting the requirements with a single-symbol pilot, \textit{i.e.}, $\tau_e=1$.  Also note the need for only $\sim 0.5$ dB of power to go from $\zeta_k = 10^{-3}$ to $\zeta_k = 10^{-5}$ for the EVT approach at the optimal solution, which is significantly smaller than the gap in multi-UE scenarios due to interference as discussed later. Furthermore, the figure shows the required power when the BS has perfect CSI knowledge, \textit{i.e.,} $\mathbf{e}_k = \mathbf{0}$, and the required power for the benchmark approach which exceeds in $\sim 1$ dB our proposed method at the optimal solution. In the following, we adopt $\tau_e = 1$ for the results corresponding to a single URLLC UE. 

\begin{figure}[t!]
    \centering
    \includegraphics[width = 1\columnwidth]{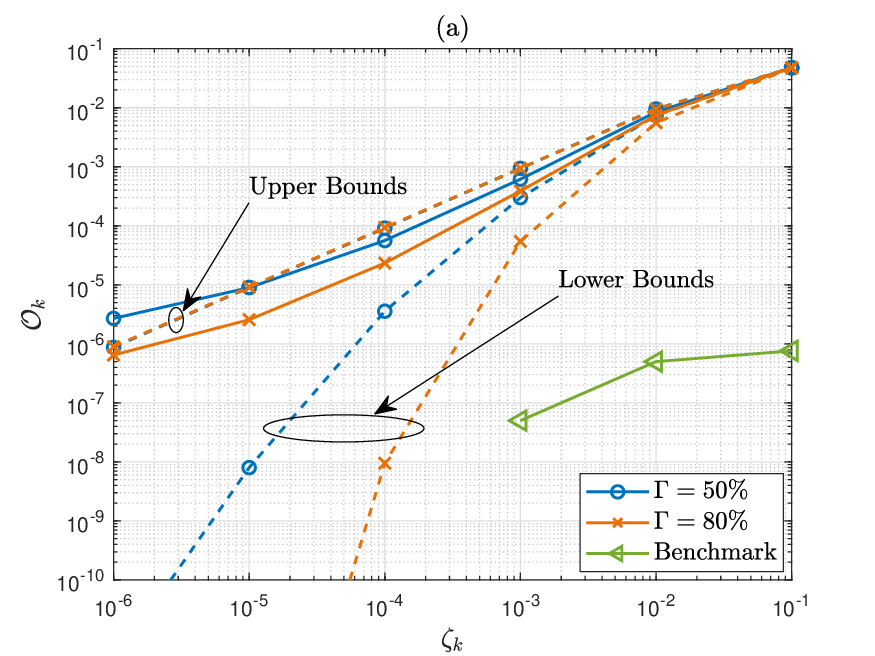}
    \includegraphics[width = 1\columnwidth]{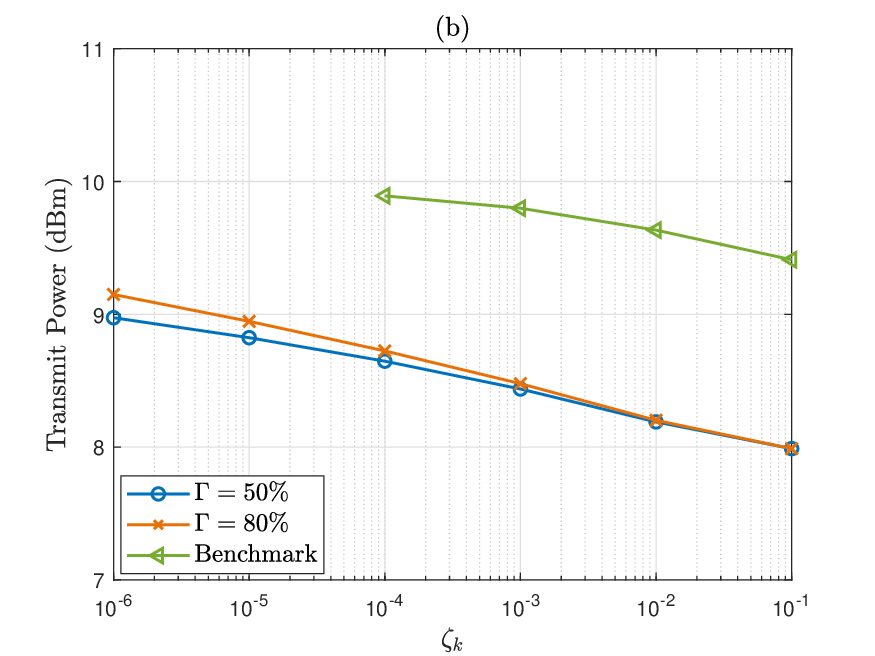}
    \caption{Outage probability bounds (a) and transmit power (b) as a function of the outage targets $\zeta_k$. We set $M=8$, $\tau_e=1$, and $N=10^4$, and employ MRT for the EVT approach.}
    \label{fig:5}
\end{figure}

Fig. \ref{fig:5} (a) shows the outage bounds as a function of the outage targets. Interestingly, the obtained upper bound obeys $\mathcal{O}_{k,\scaleto{UB}{4pt}}\approx\zeta_k$. Notice that for stricter targets, \textit{e.g.,} $\zeta_k = 10^{-6}$, the outage probability may violate the constraint if the fitting confidence is low, \textit{e.g.,} $\Gamma = 50\%$. Fig.~ \ref{fig:5} (b) displays the transmit power required to maintain the actual outage probability inside the region delimited by the upper and lower bounds in Fig \ref{fig:5} (a). The required power increases as $\zeta_k$ decreases, which is crucial for achieving ultra-reliability. Also, notice that the gap between the transmit powers for $ \Gamma=50\%$ and $\Gamma=80\%$ increases as the reliability target becomes more stringent. It is important to highlight that the energy efficiency gains with respect to the benchmark slightly increase when relaxing the outage requirements, \textit{i.e.,} increasing $\zeta_k$, since the transmit power decreases faster for the EVT scheme.

\begin{figure}[t!]
    \centering
    \includegraphics[width = 1\columnwidth]{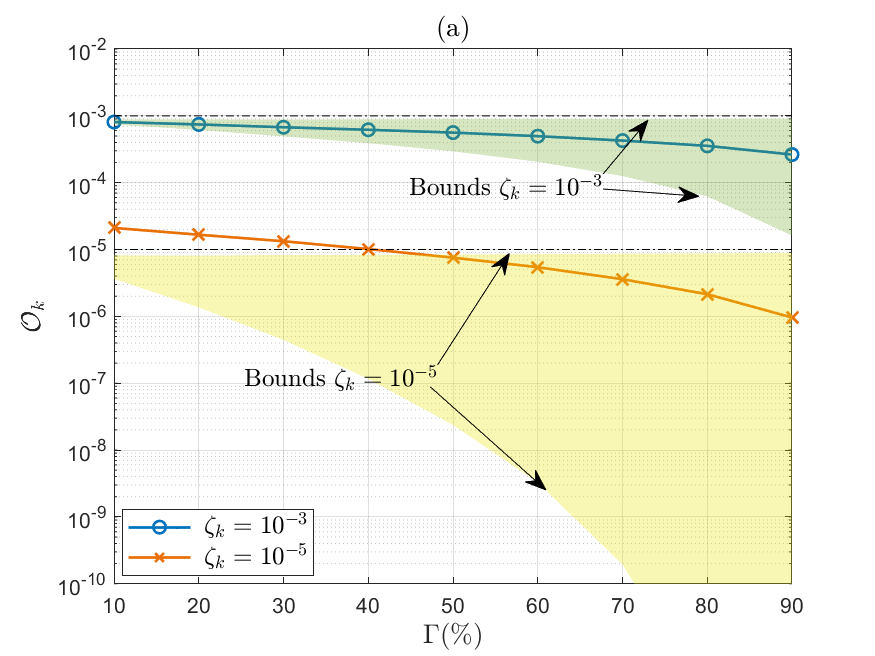}
    \includegraphics[width = 1\columnwidth]{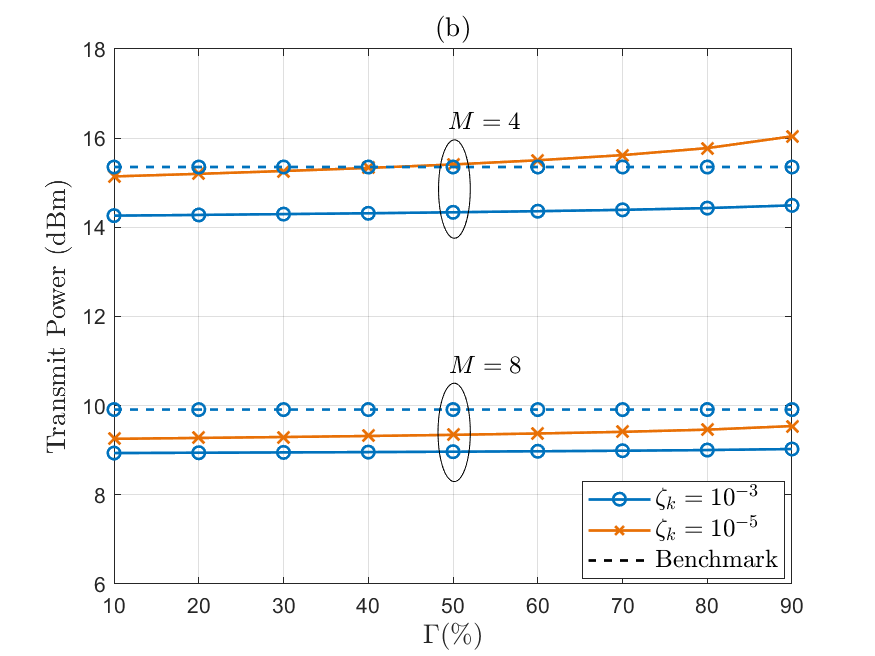}
    \caption{Outage probability (a) and transmit power (b) as a function of the confidence level $\Gamma$. We set $M \in \{4,8\}$, $\tau_e = 1$, $N = 10^4$, and employ MRT for the EVT approach. Green and yellow regions represent the outage probabilities between lower and upper bounds $\mathcal{O}_{k,LB}$ and $\mathcal{O}_{k,UB}$, respectively.}
    \label{fig:6}
\end{figure}

Fig. \ref{fig:6} (a) shows the actual outage probabilities and the outage bounds for a range of confidence levels on the GPD fitting. Notice that the target $\zeta_k = 10^{-5}$ is violated whenever $\Gamma<40\%$ which does not occur for $\zeta_k = 10^{-3}$. This suggests using a larger $\Gamma$ as the reliability requirement gets stricter. On the other hand, Fig. \ref{fig:6} (b) depicts the performance concerning required transmit power for $M = \{4,8\}$. Here, it is shown that the power gap between configurations with different confidence levels $\Gamma$ increases as $M$ decreases and the target $\zeta_k$ becomes stricter. For instance, moving from $\Gamma= 10\%$ to $\Gamma= 90\%$ with $M=8$ requires an increment of 0.09 dB and 0.286 dB for $\zeta_k=10^{-3}$ and $\zeta_k=10^{-5}$, respectively. Furthermore, with $M=4$ the increments are 0.23~dB and 0.9~dB for $\zeta_k=10^{-3}$ and $\zeta_k=10^{-5}$, respectively. This means that the fitting confidence becomes less expensive regarding power consumption as the number of antennas increases. 

\begin{figure}[t!]
    \centering
    \includegraphics[width = 1\columnwidth]{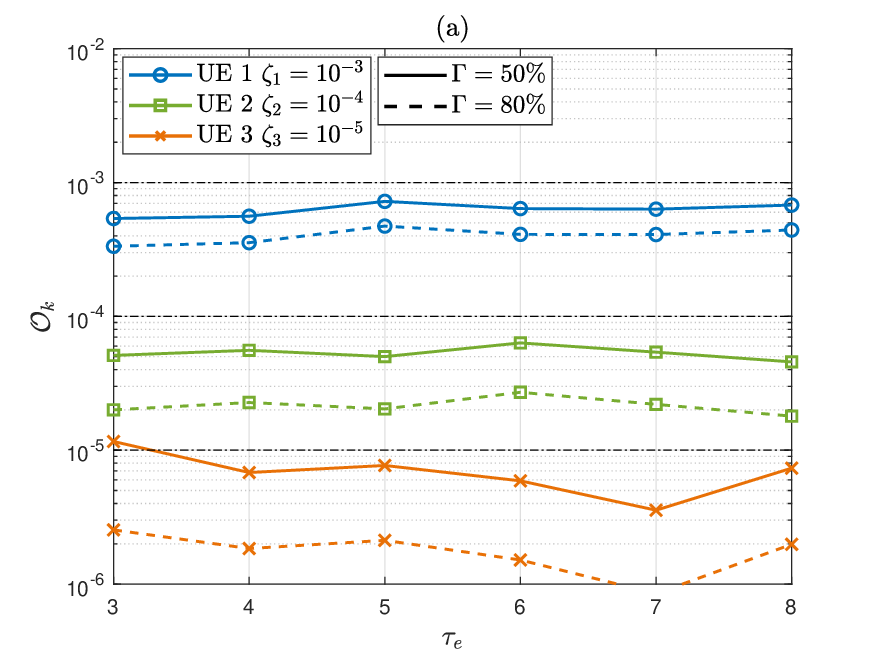}
    \includegraphics[width = 1\columnwidth]{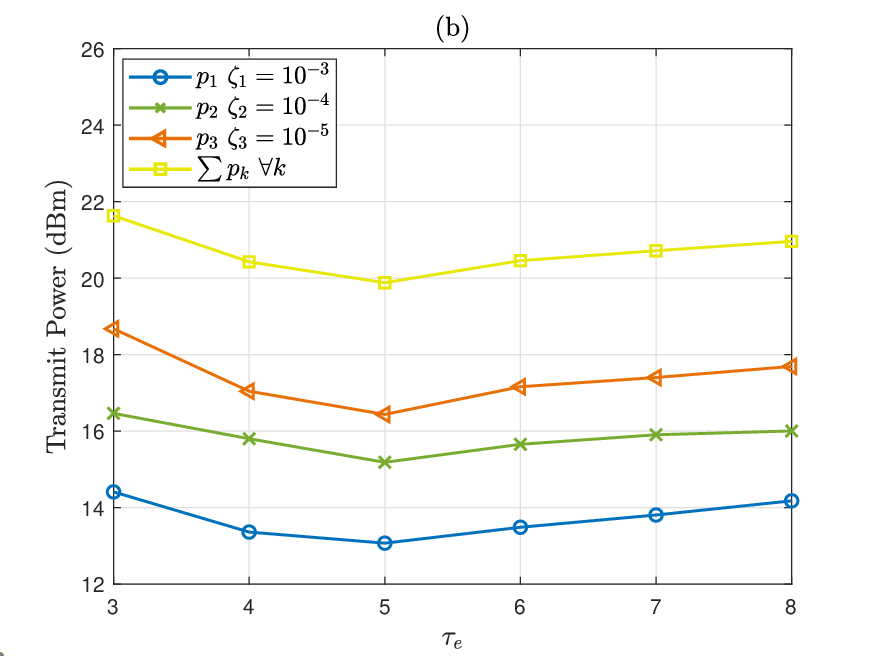}
    \caption{Outage probability (a) and transmit power (b) as a function of the pilot lengths $\tau_e$. In (b), we set $\Gamma=80\%$, $M=8$, $K=3$, and $N=10^4$, and we employ ZF precoding.}
    \label{fig:7}
\end{figure}
\subsection{Performance evaluation for multiple URLLC UEs}
Fig. \ref{fig:7} (a) and (b) show the performance of the proposed algorithm for multiple UEs in terms of outage probability and power consumption, respectively, and as a function of the pilot lengths $\tau_e$ for $\zeta_1 = 10^{-3}$, $\zeta_2 = 10^{-4}$ and $\zeta_3 = 10^{-5}$, and ZF precoding. Notice that similar to the single-UE case, the constraint $\zeta_3 = 10^{-5}$ may be violated for the case $\Gamma = 50\%$, \textit{e.g.,} for $\tau_e =3$, but all targets are guaranteed for $\Gamma =80\%$. Interestingly, in the multi-UE case, the pilot length that minimizes the total transmit power is $\tau_e = 5$. Notably, the total power is minimized with $\tau_e>K$ driven by the imperfect interference cancellation. It is worth highlighting the requirement of around 2 dB of extra power to go up or down one order of magnitude in the reliability at the optimal solution.

\begin{figure}[t!]
    \centering
    \includegraphics[width = 1\columnwidth]{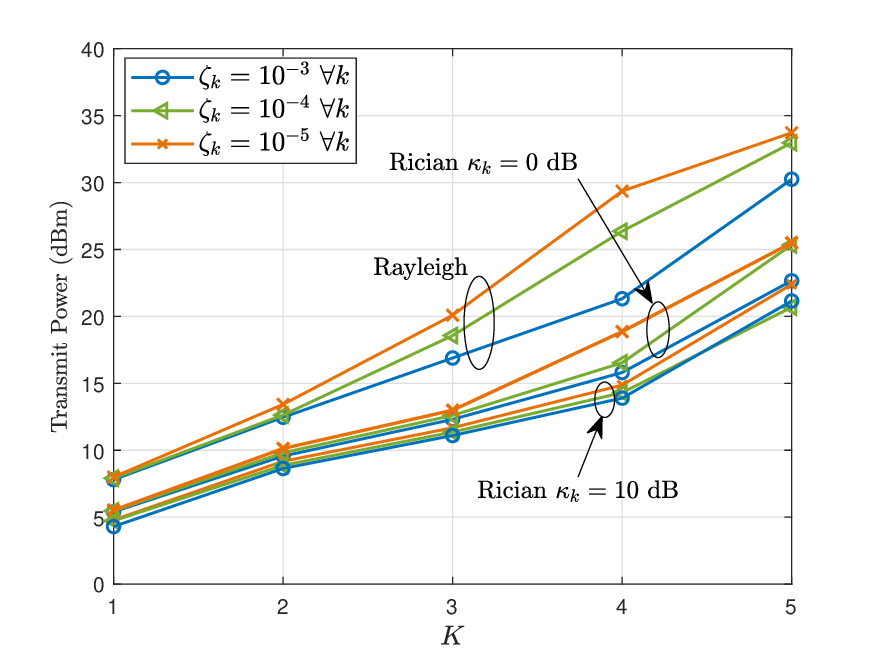}
    \caption{Transmit power as a function of the number of UEs $K$. The figure shows the performance for Rayleigh fading and Rician fading with $\kappa_k =~\{0,10\}$ dB $\forall k$, $\Gamma= 80\%$, $M = 8$, and $N = 10^4$ while employing ZF precoding.}
    \label{fig:8}
\end{figure}

Fig. \ref{fig:8} shows the transmit powers for different numbers of UEs in Rayleigh fading but also in Rician fading, \textit{i.e.,} $\mathbf{h}_k = \sqrt{\frac{\kappa_k}{\kappa_k+1}}\mathbf{h}_{k,\scaleto{LOS}{4pt}} + \sqrt{\frac{1}{\kappa_k+1}}\mathbf{h}_{k,\scaleto{NLOS}{4pt}}$ where the first and second component represent the line-of-sight (LOS) and scattering non-LOS propagation components, respectively. Moreover, $\kappa_k$ depicts the Rician factor,  $\mathbf{h}_{k,\scaleto{NLOS}{4pt}}\sim \mathcal{CN}(\mathbf{0},\mathbf{R}_k)$,  $\mathbf{h}_{k,\scaleto{LOS}{4pt}} = [1, e^{\mathbb{i} \theta_1},..., e^{\mathbb{i} \theta_{M-1}}]$ where $\theta_m$ represents the phase shift of the signal with respect of the first antenna element and $\kappa_k$ depicts the LOS factor of UE $k$. Notice that the gap in the transmit power for different outage targets increases with the number of UEs $K$. This is because the interference grows as the reliability target increases due to the increment in the required transmit power. Also, note that as the number of UEs increases, the power difference between different channel models increases for any outage target. For instance, the power difference is around 3 dB in a single-UE scenario,  while the difference is larger than 7 dB in a network with five UEs when comparing Rayleigh fading and Rician fading ($\kappa_k=0$ dB).

\section{Conclusions}\label{section_5}
 This work considered a minimum-power precoding design problem for serving multiple UEs in the DL with imperfect CSI and URLLC constraints. We proposed a solving algorithm that exploits CSI estimation error information and state-of-the-art precoding schemes such as MRT and ZF precoding. Moreover, we used the EVT framework to capture outage events that arise with low probability. Precisely, we fit data obtained from artificially-generated SINR values to the GPD with different confidence levels to model rare events in the tail of the distribution. We evaluated the performance of the presented approach through simulations and compared it with a worst-case robust precoding method in the literature. We showed that the proposed method outperforms the benchmark approach and that there is an optimal pilot length that minimizes the transmit power.  The confidence level influences the latter when fitting the data to the GPD. 

\bibliographystyle{IEEEtran}
\bibliography{IEEEabrv,bibliography}
\end{document}